\documentclass{sigchi}

\usepackage[utf8]{inputenc}
\title{Organic Visualization of Document Evolution}

\CopyrightYear{2018}
\setcopyright{acmlicensed}
\doi{10.475/123_4}
\isbn{123-4567-24-567/08/06}
\conferenceinfo{CHI'16,}{May 07--12, 2016, San Jose, CA, USA}
\acmPrice{\$15.00}
\conferenceinfo{WOODSTOCK}{'97 El Paso, Texas USA}

\numberofauthors{3}
\author{
\alignauthor
Ignacio Perez-Messina\\
\affaddr{Computer Science Dept.}\\
\affaddr{Universidad de Chile}\\
\affaddr{Santiago, Chile}\\
\email{iperez@dcc.uchile.cl}
\alignauthor
Claudio Gutierrez\\
\affaddr{Computer Science Dept.}\\
\affaddr{Universidad de Chile}\\
\affaddr{Santiago, Chile}\\
\email{cgutierr@dcc.uchile.cl}
\alignauthor Eduardo Graells-Garrido\\
\affaddr{Data Science Institute}\\
\affaddr{Universidad del Desarrollo}\\
\affaddr{Santiago, Chile}\\
\email{egraells@udd.cl}
}

\toappear{This is a pre-print version. The final version will be available in the Proceedings of the 23rd ACM Conference on Intelligent User Interfaces.}

\usepackage{balance}
\usepackage{txfonts}
\usepackage{color}
\usepackage{booktabs}
\usepackage{textcomp}
\usepackage{microtype}
\usepackage{ccicons}

\begin{document}

\maketitle

\begin{abstract}
Recent availability of data of writing processes at keystroke-granularity  has enabled research on the evolution of document writing. A natural step is to develop systems that can actually show this data and make it understandable. Here we propose a data structure that captures a document’s fine-grained history and an organic visualization that serves as an interface to it. We evaluate a proof-of-concept implementation of the system through a pilot study with documents written by students at a public university. Our results are promising and reveal facets such as general strategies adopted, local edition density and  hierarchical structure of the final text. 
\end{abstract}

\category{H.5.m.}{Information Interfaces and Presentation
  (e.g. HCI)}{Miscellaneous} 
  %\category{See \url{http://acm.org/about/class/1998/} for the full list of ACM classifiers. This section is required.}{}{}

\section{Introduction}\label{introduction}
Writing is an old and relevant human skill whose standard product is  text. Due to the widespread use of information technologies, most text available today is stored in digital platforms, such as the Web. Although static in their final form,  documents --collaborative ones in particular-- are works-in-progress, meaning they are still subject to their writing process.
Currently, we do not fully understand this everyday phenomenon, even though the way we write is tied to how we learn and structure our knowledge \cite{emig1977writing}. This may be in part because, as stated by Grésillon and Perrin in the Handbook of Writing and Text Production, ``The written (the product) aims at overcoming the writing (the process)''~\cite{gresillon2014methodology}, which means that the better the quality of a text, the more work was spent obscuring and deleting the traces of its own development. 

This lack of understanding, and of research on this topic, can be explained by the fact that data was not easily available until recently. But today, Web services such as Google Drive keep records of document changes at keystroke-level, so as data has become widely available, new avenues of research open. However, to the extent of our knowledge, this data has not yet been used to better understand the process of writing, in order, for example, to make writing easier and more effective, to help teaching, etc. In this paper we show that visualizing this data could be an important step towards understanding it.

There is a vast work on text visualization available \cite{kucher2015text}. There are two main research directions: on the one hand,  visualizations to analyze text are available, but they focus on the finished product; on the other hand, the systems that aim at the evolution of documents do so at coarse versioned text. These latter works focus on collaboration (\textit{e.g.}, Wikipedia content) and are not suited for research on individual writing.

In this paper we propose an interactive visualization method in the largely unexplored field of fine-grained text production data. Built upon organic information design guidelines, the proposed visualization shows the whole fine-grained history of a document in one image and displays its development in time with animation. It also can provide access to its textual content interactively, through which a naturally occurring segmentation of the text can be produced. By allowing complex behavior of the production of the text to visually emerge, it fosters exploration of its structure and evolution through time.
 
We evaluated our interactive visualization through a pilot study, where we visualized and analyzed documents written by engineering students. The results show different characteristics of the writing process that emerged from the visualization: general strategies adopted, local edition density, and in some cases, hierarchical structure of the final text. These results evoke interesting applications for our proposed system in fields
 including reviewing, writing teaching, assessing the depth of knowledge, among other areas. We conclude that the study and visualization of fine-grained text data enables a deep understanding of text, as it permits to augment the final product with the trace of the decisions performed during its production. 

\section{Background}\label{background}

Currently most user interfaces for text visualization focus on finished text, \textit{i.e.}, the final product of the writing process of an agent, or a corpus of such products (visual text summarization \cite{hoque2015convisit}, topic modeling \cite{sievert2014ldavis}, mapping content structure \cite{van2009mapping}, recommender systems \cite{graells2016data}, among others (see a survey \cite{vsilic2010visualization}). 

There is little work on visualization of the process itself that generates a document, particularly in the case of the human \emph{writing} process \cite{gresillon2014methodology}. 
The challenge here is to understand the structure and evolution of text according to several updates, each of which may add new content and/or delete prior content. Currently, there are two prominent sources for this kind of visual interaction: collaborative user-generated content and individual research-generated.

Regarding collaborative content, the most common data source is Wikipedia, with tools like History Flow \cite{viegas2004studying}, that visualizes the \emph{revision history} of Wikipedia articles, and the Notabilia project, which visualizes collective deliberation \cite{taraborelli2010beyond}. DocuViz \cite{wang2015docuviz} applied the History Flow approach to collaborative documents in Google Docs and Kim et al. \cite{kim2012cumulative} proposed using only document \textit{deltas} in this same line of visualization.

There are studies that focus on individual writing process~\cite{latif2008state}. Perrin and Wildi developed a statistical method to infer writing phases using cursor movement data~\cite{perrin200928}. Caporossi and Leblay~\cite{caporossi2011online} showed a graph-based visualization of the writing of a paragraph with data from ScriptLog (a keystroke logging program), where nodes represent operations; and edges their topological and temporal relations. 

Evolution of single documents has thus been researched either from a collaborative, large scale perspective using coarse data, or from an individual, fine-grained one but only at very small scale. To the best of our knowledge, there is no visualization in between that encompasses these dimensions as a whole, therefore uniting writing process research. The system we describe next aims at filling this gap.

\section{System Design}\label{system-design}

We follow the ecological \cite{wise1999ecological} and organic information design \cite{fry2000organic} approaches to create a natural-looking structure of interdependent units. We implement our prototype using the Processing language~\cite{fry2007processing}, as it's commonly used for organic visualization systems, and using data from Google Docs. Our approach is a departure from the linear, bar chart-style schema found on most of current work and aims at a similar change in the understanding of a document: not as something linear, static, but rather emergent and dynamic, but also irreversible, meaning that nothing is really deleted but submerged.

Here we describe the different stages of the pipeline needed to arrive to such depiction: the definition of text operations; the data structure holding those operations; and then the visual design that depicts the data structure.

\subsection{Representation of Document Dynamics}\label{document-dynamics}
We define a document as a chain of atomic (distinguishable in time) operations (insertions and deletions). As in Perrin’s $S$-notation \cite{perrin2003progression}, we group adjacent operations in such a way that no voluntary change in cursor position takes place between any two of them.
This results in condensed operations we call \emph{(linear keystroke) bursts}, which are more coherent and significant than single keystrokes because insertions that were immediately deleted are lost, such as correction of typographical errors (which correspond to low-level information in the writing process \cite{gresillon2014methodology}).
Finally, we reorder bursts spatially rather than temporally, as pieces of a puzzle that join one to another by the structural points we call \emph{Places of Insertion} (POIs), which are the points between characters and elements in a document where the blinking cursor can be at.

\begin{figure}[tb]
\centering
\includegraphics[width=\linewidth]{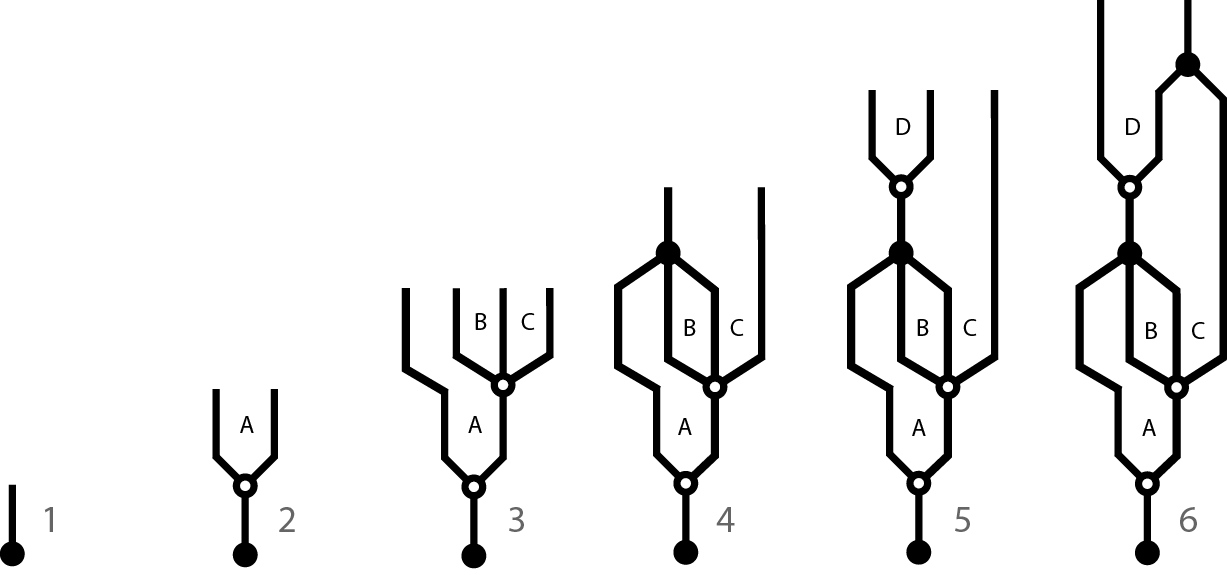}
\caption{Progression of a document's data structure. Elements: Deletion nodes (black circles), insertion nodes (white circles), edges pointing to a ``null'' node are free POIs. Edge direction is ``upward.'' (1) Empty document. (2) Insertion of ``A.'' (3) Insertion of ``BC'' at position 2, resulting in string ``ABC.'' (4) Simultaneous deletion of ``AB.'' (5) Insertion of ``D'' before ``C,'' resulting in string ``DC.'' (6) Deletion of ``C.'' Final document contains only string ``D.'' \label{fig:dag}}
\end{figure}

\subsection{Data Structure for Text
Evolution}\label{data-structure-for-text-evolution}

We store the operations and their spatial relations as a Directed Acyclic Graph, where nodes represent operations, edges are topological relations between operations and their direction follow the arrow of time (see Fig. \ref{fig:dag}). Each edge points initially to a ``null'' node, meaning a free POI. An empty document in its original state maps to a root node which contains the time of the file's creation and a single edge which stands for its only POI. At this point, only an insertion can take place (as a deletion needs more than one POI), so the next step is the addition of a new operation node containing the inserted string at the end of the root edge, from which $n+1$ edges emerge, where $n$ is the number of characters inserted, creating new POIs from the original one. This process goes on recursively, always maintaining a tree structure, but this changes when we start considering the critical aspect of deletion in the document. 
A deletion is a node that bundles together $m+1$ adjacent edges, where $m$ is the number of characters deleted, back into a single place of operation. Note that this radically changes the original insertions-only tree structure, since a deletion may encompass many levels of the hierarchy.

\subsection{Visual Representation}\label{visual-representation}
We used a glyph-based approach to visualize the aforementioned data structure, where glyphs act as interdependent units and build upon each other. Intuitively, an insertion ``opens up'' space in the document, by splitting one POI into many, while deletions ``close'' it, by joining many POIs back into one. The glyph designed to represent insertion nodes is, therefore, a stylized multiplexer. Deletion nodes, on the other hand, do not have their own glyph but retroactively affect insertion glyphs. Figure \ref{fig:glyph-schematics} illustrates this.

Seeing the visualization as a mapping from the data structure to the visual space, the rules that define this mapping are:

\begin{figure}[tb]
\centering
\includegraphics[scale=.1]{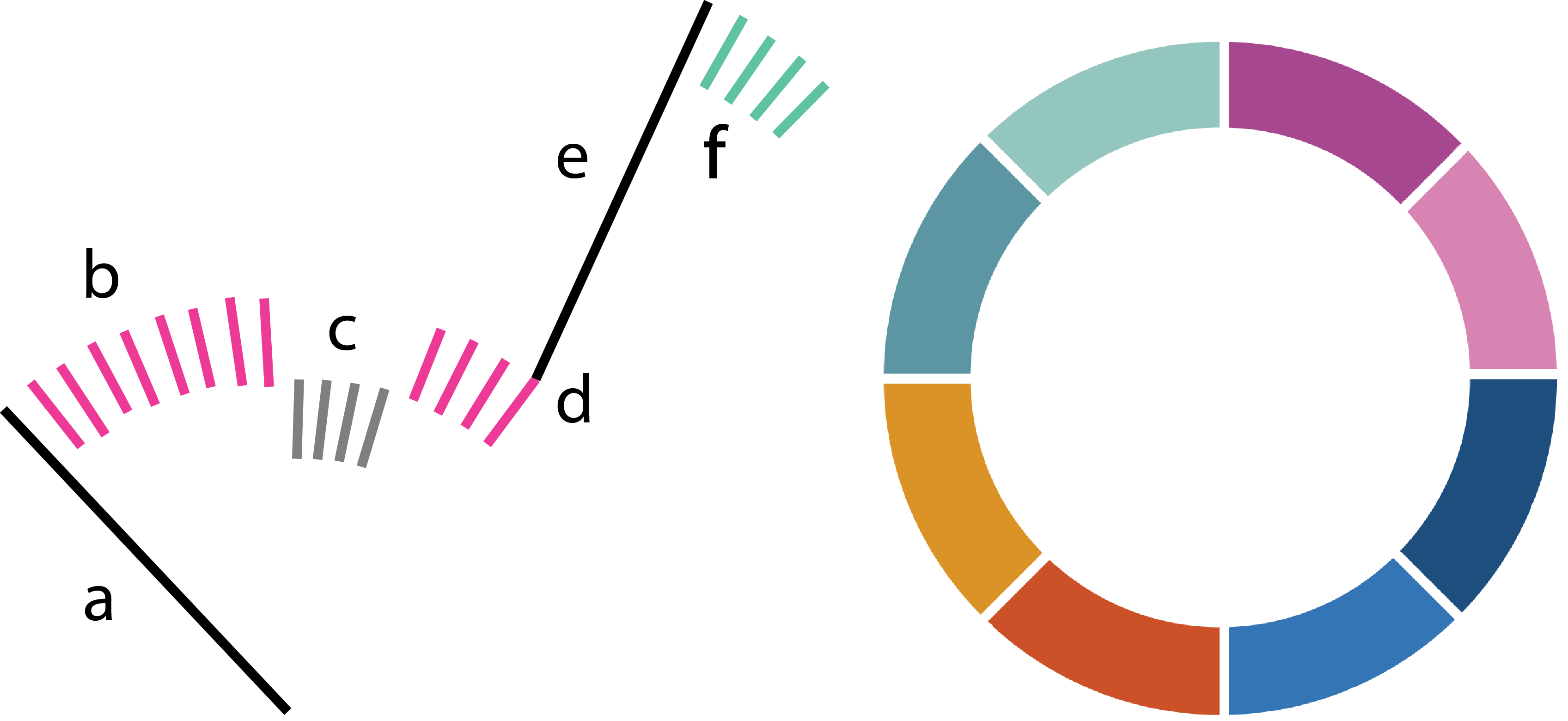}
\caption{Glyph scheme of two related nodes (left) and cyclic color palette (right). A glyph is composed by an arc (b, f), which is composed by the node's out-edges (making its length proportional to the characters inserted), and a support line (a, e). When a string is deleted, the correspondent part of the arc loses opacity and falls toward the center (c). Children nodes are placed as coming out from the POI they originated from (d). In this example, an insertion of size 15 was followed by a deletion of size 3 at position 10, and then writing was resumed at the end of the document.\label{fig:glyph-schematics}}
\end{figure}

\begin{enumerate}
\item For each insertion node, there is one glyph that represents it and its first-level out-edges.
\item An edge leading from an insertion node to another, means that the correspondent glyphs are related, precisely the latter is placed on top of the former, at the position corresponding to its relative POI within its parent.
\item And edge leading to a deletion node changes the glyph as shown in Figure \ref{fig:glyph-schematics}. 
\end{enumerate}

\textbf{Cosmetics}. To avoid spiraling branches, a ``phototropism factor'' is applied to the growth of the tree, mimicking the plant behavior of growing upwards. Time is represented using a cyclical eight-color categorical palette (see Figure \ref{fig:glyph-schematics}, right): nodes are colored according to the session (considered here as a day of writing) in which they were added. The radius of the arc doubles in case its center angle were to surpass $\pi$.

\textbf{Interaction}. ``Phototropism'' as well as arc length-node size ratio can be dynamically manipulated to globally change the shape of the tree and improve visibility. When a glyph is selected, the textual contents of its branch (deleted and active children) are displayed on  screen in a notation similar to S-notation \cite{perrin2003progression}. Parts of the tree may also be hidden at will.

%%%%%%%%%%%%%%%%%%%%%%
\section{Pilot Study}\label{pilot-study}

We performed a pilot study, where computer science students from a public university were asked to share their documents written in Google Drive. 
In total, we obtained 60 documents, of different lengths (from a few paragraphs to full-length articles), and purposes (though they were all course assignments). Most of them were ruled out before visual analysis due to incompletion or not showing enough complexity. We selected five documents to show here due to their complementariness (see Table \ref{documents}). For each one, we identified the visualization's branching structure, which leads to a hierarchic segmentation of the tree. Then, we inspected each branch's content, and identified which part of the document corresponded to the branch. We also took note of branch length and breadth, and important deletions, which we interpreted in the context of each document.

\begin{table}[tb]
\centering
\caption{Description of document case studies.}
\label{documents}
\begin{tabular}{llll}
\toprule
Doc      & Description                  & Words & Operations \\ \midrule
A        & Two-item summary        & 312   & 1307       \\
B        & Three-question assignment    & 1567  & 7136       \\
C        & One-question assignment   & 657   & 3015       \\
D        & Unstructured essay & 5242  & 15411      \\
E		 & Structured assignment & 1135  & 4350 \\
\bottomrule
\end{tabular}
\end{table}

\begin{figure}[t]
\centering
\includegraphics[width=\linewidth]{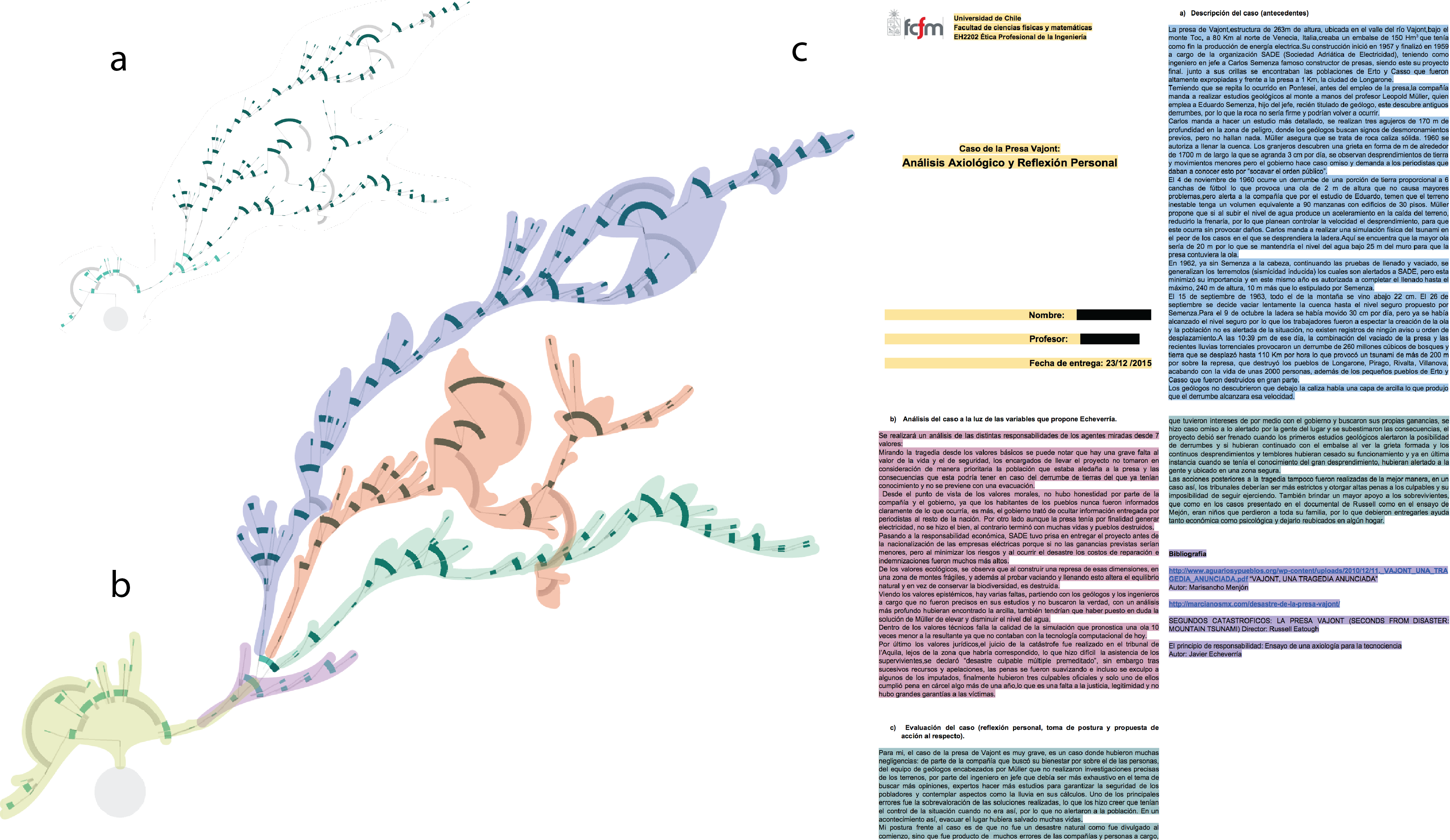}
\caption{Analysis of document B. From the raw visualization (a), we identified its branching structure (b; colors were manually added). Inspecting each branch's content produces a correspondent segmentation of the text (c; highlights were manually added to match correspondent colors in b). This shows that the branching structure of the tree is the same as the hierarchic structure of the document: the cover title splits into two, the bibliography and the body, which splits also in three sections. This in turn means that the writing process of this document followed its prescribed structure.
\label{fig:star-example}}
\end{figure}

\begin{figure}[t]
\centering
\includegraphics[width=\linewidth]{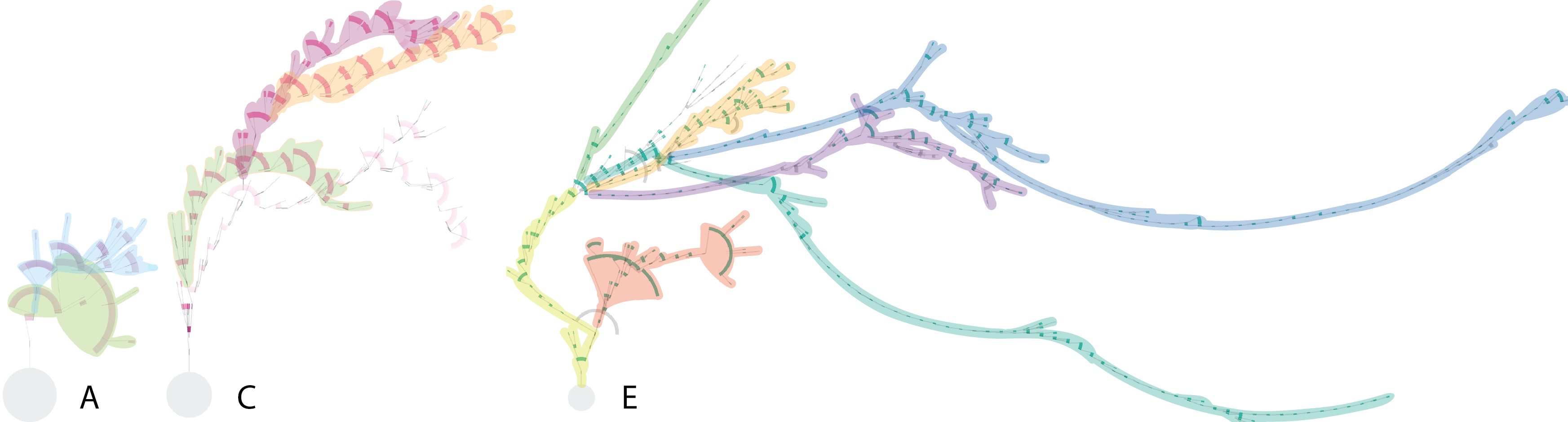}
\caption{Visualization of documents A, C and E. Note that the branching structure of each tree was manually highlighted to account for its corresponding parts on each document. A shows two branches, corresponding to its two paragraphs. C has a richer structure, featuring four main branches, one of which (the faint-looking one) was completely deleted (we say it's ``dead'') and from which the other branches arise. E is more complex, each color mapping to a section of the text, same as case B. These cases show different writing strategies: almost linear (A), draft and rewrite (C), and hierarchically structured (E).\label{fig:case_studies}}
\end{figure}

There are three dimensions of document evolution that, according to the analysis, are well captured by our system:

\begin{figure}[t]
\centering
\includegraphics[width=\linewidth]{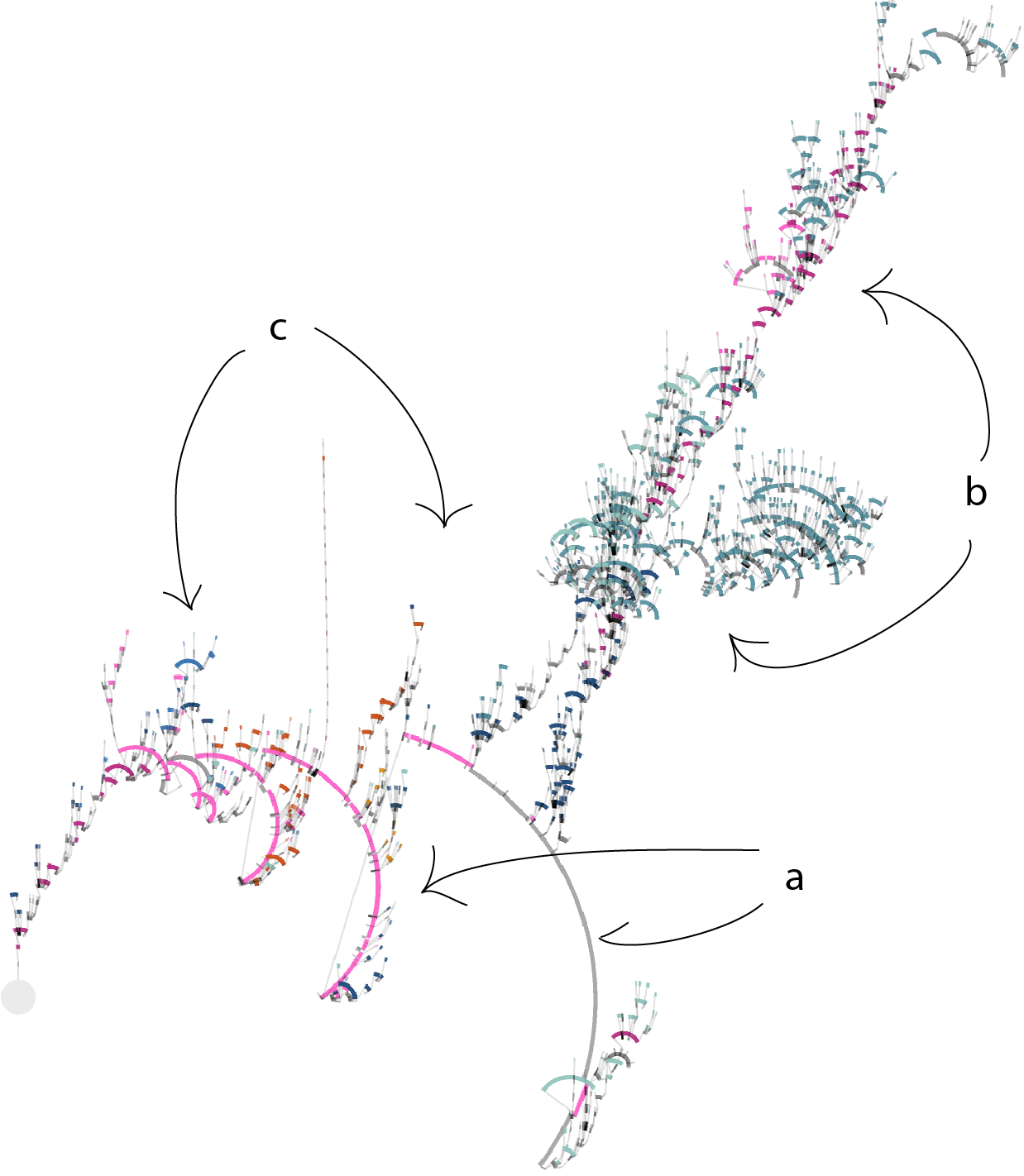}
\caption{Analysis of D. Very long arcs (a) denote copy/pastes. Branches extending from these arcs (b; note that the largest one was almost completely deleted) are their rewriting and have the highest density of changes. Little branches of different colors (c) are later additions. \label{fig:41}}
\end{figure}

\begin{enumerate}
\def\labelenumi{\arabic{enumi}.}
\item
  \emph{The internal organization of the text and its hierarchic structure} (Fig. \ref{fig:star-example}).
  We observed that branches of a tree mostly correspond to hierarchical structure of the text. In Cases A and C, branches match paragraph divisions, as they have no other hierarchical level. Cases B and E have a typical hierarchical organization (cover information, sections, and bibliography) which is perfectly matched by the relations of the correspondent branches. Case D has also no more structure than paragraph-level as can be intuited by its ``one big branch'' appearance.
\item 
  \emph{Some patterns and strategies adopted by users} (Fig. \ref{fig:case_studies}). The structure of the tree reveals also the strategies used to write the document. Cases B and E, for example, show a well-defined hierarchical structure, meaning its writing bore the final structure in mind from the beginning, something that can be expected in a course assignment.
Case B shows a draft that was rewritten and erased, while A was written almost linearly, without important deletions.
\item
  \emph{The amount of work put into the document and its different parts} . This dimension emerges from color heterogeneity and glyph density of a branch. Case studies A, B, C and E have branches of only one color, meaning they were introduced during one session with no later rewriting, whereas D (Fig. \ref{fig:41}) has branches showing many appendices of different colors, meaning they were reread and edited in posterior sessions. Moreover, the highest edition density is concentrated around the deletion of a large piece of text that was pasted from another document.

\end{enumerate}

In summary, we observed that the system captures important components of the writing process. Next we discuss the implications and future work due to these findings. 

\section{Discussion and Conclusions}\label{discussion}
Our results shed light on the dynamic origins of text and the structures underlying the process of writing. These findings could be useful in education (\emph{e.g.}, evaluation and assessment of learning), work (\emph{e.g.}, matching thinking structure to teams, which could be used in hiring processes), and natural language processing (\emph{e.g.}, by including human-writing processes into automated document generation, or document summarization).
A direct application of our system is a real-time writing-aid in document writing tools, which returned to the document its heterogeneity, for example, showing the relative age of parts of text, their need for update and the thread they belong to.

\textbf{Scope and Future Work}. A rightful critique is that, owing to its lack of a different glyph for deletion nodes, the visualization captures only a subset of the data structure, \emph{i.e.}, it is only a spanning tree of the whole graph, which leads to the non-uniqueness of a document’s representation: a design fault because it forces a degree of freedom not present in the data~\cite{kindlmann2014algebraic}. Future work, then, should include the design of deletion nodes so that they play a structural role. Also, branch overlapping is a major problem, which currently makes it impossible to analyze larger documents. A solution for this would be the implementation of glyph space-awareness, and interactive expansion. Finally, the pilot study showed that coloring a tree by its branching structure is a necessary step for analysis, so an interesting intelligent feature would be the automation of this segmentation and highlighting, linking it to the final document.

\textbf{Conclusions}.
We have presented a novel visualization design for document evolution which combines an operational view of the document with an organic visual scheme, and have shown that it renders visible some complex behavior in writing. 
It can be used, for example, to get an overview of the whole of a document’s history in a single image, which is enough to give an idea of the amount of work put into it and the general strategy adopted. Examples of such strategies are rewriting from a draft, writing with a structure in mind, one- \emph{vs.} many-session writing, etc. These features are something that, for a single session or single user document and at this level of granularity, to the best of
our knowledge, available systems cannot provide. Also, with its interactive functions, the system can be used to produce a segmentation of a document, which in some cases coincides with its hierarchical structure, but in any case is a naturally occurring segmentation which follows the thread of thought of the user.
We present this approach and system to provide an integration of computer-aided writing research by proposing a clear focus on the document as a well-defined temporal object.approach and system Topic segmentation should not be abstracted from a document’s history when possible, and this approach proves a fair candidate for segmenting a document through its own writing history.

\balance{}

\bibliographystyle{SIGCHI-Reference-Format}
\bibliography{references}


\begin{thebibliography}{00}

%%% the Web address.

\ifx \showCODEN    \undefined \def \showCODEN     #1{\unskip}     \fi
\ifx \showDOI      \undefined \def \showDOI       #1{{\tt DOI:}\penalty0{#1}\ }
  \fi
\ifx \showISBNx    \undefined \def \showISBNx     #1{\unskip}     \fi
\ifx \showISBNxiii \undefined \def \showISBNxiii  #1{\unskip}     \fi
\ifx \showISSN     \undefined \def \showISSN      #1{\unskip}     \fi
\ifx \showLCCN     \undefined \def \showLCCN      #1{\unskip}     \fi
\ifx \shownote     \undefined \def \shownote      #1{#1}          \fi
\ifx \showarticletitle \undefined \def \showarticletitle #1{#1}   \fi
\ifx \showURL      \undefined \def \showURL       #1{#1}          \fi

\bibitem{caporossi2011online}
{Gilles Caporossi} {and} {Christophe Leblay}. 2011.
\newblock \showarticletitle{Online writing data representation: a graph theory
  approach}. In {\em International Symposium on Intelligent Data Analysis}.
  Springer, 80--89.
\newblock


\bibitem{emig1977writing}
{Janet Emig}. 1977.
\newblock \showarticletitle{Writing as a mode of learning}.
\newblock {\em College composition and communication\/} {28}, 2 (1977),
  122--128.
\newblock


\bibitem{fry2007processing}
{Ben Fry}. 2007.
\newblock {\em A Processing: Programming Handbook for Visual Designers and
  Artists}.
\newblock MIT Press.
\newblock


\bibitem{fry2000organic}
{Benjamin~Jotham Fry}. 2000.
\newblock {\em Organic information design}.
\newblock Ph.D. Dissertation. Massachusetts Institute of Technology.
\newblock


\bibitem{graells2016data}
{Eduardo Graells-Garrido}, {Mounia Lalmas}, {and} {Ricardo Baeza-Yates}. 2016.
\newblock \showarticletitle{Data portraits and intermediary topics: Encouraging
  exploration of politically diverse profiles}. In {\em Proceedings of the 21st
  International Conference on Intelligent User Interfaces}. ACM, 228--240.
\newblock


\bibitem{gresillon2014methodology}
{Almuth Gr{\'e}sillon} {and} {Daniel Perrin}. 2014.
\newblock \showarticletitle{Methodology: From speaking about writing to
  tracking text production}.
\newblock {\em Handbook of writing and text production\/}  {10} (2014),
  79--111.
\newblock


\bibitem{hoque2015convisit}
{Enamul Hoque} {and} {Giuseppe Carenini}. 2015.
\newblock \showarticletitle{Convisit: Interactive topic modeling for exploring
  asynchronous online conversations}. In {\em Proceedings of the 20th
  International Conference on Intelligent User Interfaces}. ACM, 169--180.
\newblock


\bibitem{kim2012cumulative}
{Seungyeon Kim}, {Joshua~V Dillon}, {and} {Guy Lebanon}. 2012.
\newblock \showarticletitle{Cumulative Revision Map}.
\newblock {\em arXiv preprint arXiv:1205.3205\/} (2012).
\newblock


\bibitem{kindlmann2014algebraic}
{Gordon Kindlmann} {and} {Carlos Scheidegger}. 2014.
\newblock \showarticletitle{An algebraic process for visualization design}.
\newblock {\em IEEE transactions on visualization and computer graphics\/}
  {20}, 12 (2014), 2181--2190.
\newblock


\bibitem{kucher2015text}
{Kostiantyn Kucher} {and} {Andreas Kerren}. 2015.
\newblock \showarticletitle{Text visualization techniques: Taxonomy, visual
  survey, and community insights}. In {\em Visualization Symposium
  (PacificVis), 2015 IEEE Pacific}. IEEE, 117--121.
\newblock


\bibitem{latif2008state}
{Muhammad M~Abdel Latif}. 2008.
\newblock \showarticletitle{A state-of-the-art review of the real-time
  computer-aided study of the writing process}.
\newblock {\em International Journal of English Studies\/} {8}, 1 (2008),
  29--50.
\newblock


\bibitem{perrin2003progression}
{Daniel Perrin}. 2003.
\newblock \showarticletitle{Progression analysis (PA): Investigating writing
  strategies at the workplace}.
\newblock {\em Journal of Pragmatics\/} {35}, 6 (2003), 907--921.
\newblock


\bibitem{perrin200928}
{Daniel Perrin} {and} {Marc Wildi}. 2009.
\newblock \showarticletitle{28 Statistical modeling of writing processes}.
\newblock {\em Traditions of writing research\/} (2009), 378.
\newblock


\bibitem{sievert2014ldavis}
{Carson Sievert} {and} {Kenneth~E Shirley}. 2014.
\newblock \showarticletitle{LDAvis: A method for visualizing and interpreting
  topics}. In {\em Proceedings of the workshop on interactive language
  learning, visualization, and interfaces}. 63--70.
\newblock


\bibitem{vsilic2010visualization}
{Artur {\v{S}}ili{\'c}} {and} {Bojana~Dalbelo Ba{\v{s}}i{\'c}}. 2010.
\newblock \showarticletitle{Visualization of text streams: A survey}. In {\em
  International Conference on Knowledge-Based and Intelligent Information and
  Engineering Systems}. Springer, 31--43.
\newblock


\bibitem{taraborelli2010beyond}
{Dario Taraborelli} {and} {Giovanni~Luca Ciampaglia}. 2010.
\newblock \showarticletitle{Beyond notability. Collective deliberation on
  content inclusion in Wikipedia}. In {\em Self-Adaptive and Self-Organizing
  Systems Workshop (SASOW), 2010 Fourth IEEE International Conference on}.
  IEEE, 122--125.
\newblock


\bibitem{van2009mapping}
{Frank Van~Ham}, {Martin Wattenberg}, {and} {Fernanda~B Vi{\'e}gas}. 2009.
\newblock \showarticletitle{Mapping text with phrase nets}.
\newblock {\em IEEE transactions on visualization and computer graphics\/}
  {15}, 6 (2009).
\newblock


\bibitem{viegas2004studying}
{Fernanda~B Vi{\'e}gas}, {Martin Wattenberg}, {and} {Kushal Dave}. 2004.
\newblock \showarticletitle{Studying cooperation and conflict between authors
  with history flow visualizations}. In {\em Proceedings of the SIGCHI
  conference on Human factors in computing systems}. ACM, 575--582.
\newblock


\bibitem{wang2015docuviz}
{Dakuo Wang}, {Judith~S Olson}, {Jingwen Zhang}, {Trung Nguyen}, {and} {Gary~M
  Olson}. 2015.
\newblock \showarticletitle{DocuViz: visualizing collaborative writing}. In
  {\em Proceedings of the 33rd Annual ACM Conference on Human Factors in
  Computing Systems}. ACM, 1865--1874.
\newblock


\bibitem{wise1999ecological}
{James~A Wise}. 1999.
\newblock \showarticletitle{The ecological approach to text visualization}.
\newblock {\em Journal of the Association for Information Science and
  Technology\/} {50}, 13 (1999), 1224.
\newblock


\end{thebibliography}

\end{document}